\documentclass[twocolumn,showpacs,prl,amsmath,amssymb]{revtex4}

\usepackage{graphicx}
\usepackage{dcolumn}
\usepackage{bm}
\usepackage{pdfpages}

\newcommand{\be}{\begin{equation}}
\newcommand{\ee}{\end{equation}} 
\newcommand{\lb}{\label}

\newcommand{\ba}{{\bf a}}

\newcommand{\bF}{{\bf f}}
\newcommand{\bk}{{\bf k}}

\newcommand{\br}{{\bf r}}
\newcommand{\bu}{{\bf u}}

\newcommand{\bx}{{\bf x}}

\newcommand{\bJ}{{\bf J}}

\newcommand{\OL}{\overline}

\newcommand{\wt}{\widetilde}

\newcommand{\bomega}{{\mbox{\boldmath $\omega$}}}

\newcommand{\grad}{{\mbox{\boldmath $\nabla$}}}
\newcommand{\bdot}{{\mbox{\boldmath $\cdot$}}}

\newcommand{\btimes}{{\mbox{\boldmath $\times$}}}

\newcommand{\boell}{{\mbox{\boldmath $\ell$}}}

\begin{document}

\preprint{APS/123-QED}

\title{Compressible Turbulence: The Cascade and its Locality}

\author{Hussein Aluie}
\affiliation{Applied Mathematics and Plasma Physics (T-5)
\& Center for Non-linear Studies,\\
 Los Alamos National Laboratory, MS-B258
Los Alamos, NM 87545, USA}

\date{\today}

\begin{abstract}
We prove that inter-scale transfer of kinetic energy in compressible turbulence 
is dominated by local interactions. In particular, our results preclude direct transfer of kinetic energy from
large-scales directly to dissipation scales, such as into shocks, in high Reynolds number turbulence
 as is commonly believed. Our assumptions on the scaling of structure functions are weak and 
 enjoy compelling empirical support.  Under a stronger assumption on pressure dilatation co-spectrum, 
 we show that 
mean kinetic and internal energy budgets statistically decouple beyond a transitional
``conversion'' range. Our analysis establishes the existence of an ensuing inertial range
over which  mean SGS kinetic energy flux becomes constant, independent of scale.
Over this inertial range, mean kinetic energy cascades locally and in a conservative fashion,
despite not being an invariant.
\end{abstract}
\pacs{47.27.-i, 47.40.-x, 47.27.E-, 47.27.Jv, 47.27.em}
\maketitle

Turbulence is a phenomenon that pervades most liquid, gas, and plasma flows 
in engineering and nature, ranging from high-speed engines, 
nuclear fusion power reactors, and spacecraft re-entry,
to star formation in molecular clouds, and supernovae. While the traditional
Richardson-Kolmogorov-Onsager picture is a successful theory of incompressible 
turbulence, all aforementioned systems are characterized by significant  compressibility effects.
We have recently laid a rigorous framework in \cite{Aluie10a}
to study scale-coupling in compressible flows and to analyze transfer 
of kinetic energy between different scales. The purpose of this Letter is to explore 
if such transfer of energy takes place through a cascade process and whether the 
cascade is scale-local.

Kolmogorov's 1941 theory of incompressible turbulence 
makes the fundamental assumption of a scale-local cascade process in which modes
all of a comparable scale $\sim \ell$ 
participate predominantly in the transfer of energy across scale $\ell$. 
If, furthermore, the cascade steps are chaotic processes
then it is expected that any ``memory'' of large-scale particulars of the system, such as 
geometry and large-scale statistics, or the specifics of microscopic dissipation,
will be ``forgotten.'' This gives rise to an inertial scale-range
over which turbulent fluctuations have universal statistics
and the flow evolves under its own internal dynamics
without \emph{direct} communication with the largest or smallest scales in the system.

Therefore, scale locality of the cascade is crucial to justify the existence of 
universal statistics and to warrant the concept of an inertial range. It is, furthermore,
necessary for the physical foundation of large-eddy simulation (LES) modelling of 
turbulence. It motivates the belief that models of subscale terms in the equations for 
large-scales can be of general utility, independent of the particulars of turbulent flows 
under study. While scale-locality in incompressible turbulence stands on firm
theoretical \cite{Kraichnan59,Eyink05} and numerical \cite{Domaradzkietal09,AluieEyink09}
grounds, no similar results exist for compressible turbulence.
In fact, there is a widespread belief especially common in the astrophysical literature which 
maintains that a ``finite portion'' of energy at a given scale must be dissipated directly into 
shocks through non-local transfer in scale (see for example \cite{McKeeOstriker07}).
Moreover, the idea of a cascade
itself is without physical basis since kinetic energy is not a global invariant of the inviscid 
dynamics. Hence, the notion of an inertial cascade-range in compressible turbulence
 remains tenuous and unsubstantiated.

In this Letter, we prove under modest assumptions that transfer of kinetic energy is indeed local
in scale. Under a stronger assumption, we will further show that kinetic energy cascades
conservatively despite not being an invariant. We reach these results by a direct analysis of the 
compressible Navier Stokes equations, without use of any closure approximation. 
The equations are those of continuity and momentum:
\begin{eqnarray} 
\partial_t \rho + \grad\bdot(\rho \bu) = 0, ~~~~~~~~~~~~~~~~~~~~~~~~~~~~~~~~~~~~~~~~~~~~~~~~\lb{continuity} \\
\partial_t (\rho \bu) + \grad\bdot(\rho \bu \bu) 
= -\grad P +   \mu\grad\bdot(\grad \bu +\frac{1}{3}\grad\bdot \bu{\bf I})  + \rho \bF, \lb{momentum}\\
\nonumber\end{eqnarray}
and either internal or total energy, supplemented with an equation of state for the fluid.
Here, $\bF$ is an external acceleration field stirring the fluid, 
and we have assumed a constant dynamic viscosity, $\mu$. 

Our analysis is based on a coarse-graining (or filtering) approach expounded in 
\cite{Aluie10a}, in which we observed that any scale decomposition aimed
at studying inertial-range dynamics must satisfy an \emph{inviscid criterion},
{\it i.e.} it must guarantee that viscous momentum diffusion and kinetic energy dissipation are negligible
at large-scales. We proved that a Favre decomposition meets such a requirement. 
Using classically filtered fields, $\OL \ba_\ell(\bx) \equiv \int d^3\br~ G_\ell(\br) \ba(\bx+\br)$,
with kernel $G_\ell(\br) = \ell^{-3} G(\br/\ell)$ that is smooth and decays sufficiently rapidly for large $r$,
a Favre filtered field is weighted by density as 
$\wt{\ba}_\ell(\bx) \equiv \OL{\rho \ba}_\ell(\bx)/\OL\rho_\ell(\bx)$.
From coarse-grained continuity and momentum equations, we can write down
a large-scale kinetic energy budget,
\be
\partial_t \OL\rho\frac{|\wt\bu|^2}{2} + \grad\bdot\bJ_\ell 
= -\Pi_\ell -\Lambda_\ell + \OL{P}_\ell\grad\bdot\OL\bu_\ell
-D_\ell
+\epsilon^{inj}.
\lb{largeKE}\ee
Eq. (\ref{largeKE}) describes instantaneous kinetic energy evolution at every point $\bx$ in the flow
and at scales $>\ell$, for arbitrary $\ell$. Our approach, therefore, allows the simultaneous resolution 
of dynamics both in scale and in space, and admits intuitive physical interpretation of all terms.
Here $\bJ_\ell(\bx)$ is spatial transport of large-scale kinetic energy, 
$-\OL{P}\grad\bdot\OL\bu$ is large-scale
pressure dilatation, $D_\ell(\bx)$ is viscous dissipation acting
on scales $>\ell$, and $\epsilon^{inj}(\bx)$ is the energy injected due to external stirring
(see \cite{Aluie10a} for details). We have proved in \cite{Aluie10a} that $D_\ell(\bx)$
is negligible at scales $\ell\gg\ell_\mu$, where $\ell_\mu$ denotes the dissipation scale.
We have also shown that mean kinetic energy injection can be localized to the 
largest scales $L\gg\ell$ by proper stirring. Over an intermediate scale-range
$L\gg\ell\gg\ell_\mu$, the only relevant terms in eq.(\ref{largeKE}) are inertial processes.
The subgrid scale (SGS) flux terms are defined as 
\begin{eqnarray}
&\Pi_\ell(\bx)& = ~  -\OL{\rho}~ \partial_j \wt{u}_i  ~ \wt\tau(u_i,u_j)  \lb{flux1}\\
& \Lambda_\ell(\bx)& = ~  \frac{1}{\OL\rho}\partial_j\OL{P}~\OL\tau(\rho,u_j), \lb{flux2}
\end{eqnarray}
and act as sinks in (\ref{largeKE}), transferring large-scale kinetic energy 
to scales $<\ell$. Deformation work, $\Pi_\ell$, is
due to large-scale strain, $\grad\wt\bu_\ell$, acting against turbulent stress,
$\OL\rho_\ell\wt\tau_\ell(\bu,\bu) =\OL\rho(\wt{(\bu\bu)}_\ell-\wt{\bu}_\ell\wt{\bu}_\ell)$,
while baropycnal work, $\Lambda_\ell$, is due to large-scale
pressure-gradient force, $\grad \OL{P}_\ell/\OL\rho_\ell$, acting against 
turbulent mass flux, $\OL\tau_\ell(\rho,\bu)$ (see \cite{Aluie10a} for a more detailed
discussion of the physics).
We employ the notation $\OL\tau_\ell(f,g) \equiv\OL{(fg)}_\ell-\OL{f}_\ell\OL{g}_\ell$ 
for $2^{nd}$-order central moments of fields $f(\bx),g(\bx)$ \cite{Germano92}. 

There are three facts crucial for proving scale locality of inter-scale transfer.
First is the observation that deformation work, $\Pi_\ell$, and baropycnal work, $\Lambda_\ell$,
represent the only two processes capable of direct transfer of kinetic energy \emph{across} scales.
Pressure dilatation, $-\OL{P}_\ell\grad\bdot\OL\bu_\ell$, does not contain any modes at scales $<\ell$
and does not vanish in the absence of subscale fluctuations.
It, therefore, cannot participate in transferring kinetic energy directly across scales and only 
contributes to conversion of large-scale kinetic energy into internal energy. 
This observation allows us to circumvent analyzing the internal energy budget
which does not couple to large-scale kinetic energy via viscous dynamics,
as we have proved in \cite{Aluie10a}.

The remaining two parts of our proof build upon previous studies in incompressible 
hydrodynamic \cite{Eyink05} and magnetohydrodynamic \cite{AluieEyink10} turbulence,
with some technical modifications. The second ingredient we use is the fact that 
SGS kinetic energy flux across $\ell$, $\Pi_\ell+\Lambda_\ell$, can be written in terms of increments, 
$\delta f(\bx;\br)=  f(\bx+\br)-f(\bx)$,
for separation distances $|\br|<\ell$ (or some moderate multiple of $\ell$)
and do not depend on the absolute field $f(\bx)$. Baropycnal
work, $\Lambda_\ell$, can be expressed in terms of increments by noting that 
gradient fields and central moments are related to increments as
\begin{eqnarray}
\grad\OL{f}_\ell = {\cal O}[\delta f(\ell)/\ell], \,\,\,\,\,\,\,\,f'_\ell = {\cal O} [\delta f(\ell)],\nonumber\\
\OL\tau_\ell(f,g) = {\cal O}[\delta f(\ell)~\delta g(\ell)]~~~~~~~~~
\lb{incrementrelation1}\end{eqnarray}
where symbol ${\cal O}$ stands for ``same order-of-magnitude as,''
and $f'_\ell = f-\OL{f}_\ell$ is the fine-scale field.
For rigorous details, see \cite{Eyink05,Aluie10b}.
In order to express deformation work, $\Pi_\ell$, in terms of increments, 
we need the following identities which are straightforward to verify:
\begin{eqnarray}
\grad\wt{\bu} =  \grad\OL{\bu} + {\OL\rho}^{-1}\grad\OL\tau(\rho,\bu) - {\OL\rho}^{-2}\OL\tau(\rho,\bu)\grad\OL\rho,~~~~~~~~~~~~~
\lb{identity1}\\
\wt\tau(\bu,\bu) = \OL\tau(\bu,\bu) + \OL\tau(\rho,\bu,\bu)/{\OL\rho} - \OL\tau(\rho,\bu)\OL\tau(\rho,\bu)/{\OL\rho}^{2}.~~\lb{identity2}
\end{eqnarray}
We are finally able to express $\Pi_\ell$ in terms of increments using (\ref{incrementrelation1})
and two additional relations,
\begin{eqnarray}
\grad \OL\tau_\ell(f,g) = {\cal O}[\delta f(\ell)~\delta g(\ell)/\ell],\nonumber\\
\OL\tau_\ell(f,g,h) = {\cal O}[\delta f(\ell)~\delta g(\ell)~\delta h(\ell)],
\lb{incrementrelation2} \end{eqnarray}
whose rigorous details are in our longer work \cite{Aluie10b}.
The relation of $3^{rd}$-order central moments,
$\OL\tau(f,g,h)
\equiv\OL{(fgh)}_\ell
-\OL{f}_\ell\OL{\tau}_\ell(g,h) -\OL{g}_\ell\OL{\tau}_\ell(f,h) -\OL{h}_\ell\OL{\tau}_\ell(f,g) -\OL{f}_\ell\OL{g}_\ell\OL{h}_\ell
$, to increments is unpublished and due to G. L. Eyink \cite{EyinkNotes}.

Since $\Pi_\ell$ and $\Lambda_\ell$ can be expressed in terms of velocity, pressure, and density increments, it thus 
becomes sufficient to show that these increments themselves are 
scale-local. To establish this, we need the third requirement crucial for locality ---that
scaling properties of structure functions of velocity, pressure, and density increments 
are constrained by:
\begin{eqnarray}
\| \delta\bu(\br) \|_p &\sim& u_{rms} A_p (r/L)^{\sigma^u_p}, \,\,\,\,\,\,\,\,\,\,\,\, 0<\sigma^u_p<1,\lb{scaling1}\\
\| \delta P(\br) \|_p &\sim& P_{rms} B_p (r/L)^{\sigma^P_p}, \,\,\,\,\,\,~~~~\,\,\,\,\,\,\, \sigma^P_p<1,\lb{scaling2}\\
\| \delta\rho(\br) \|_p &\le& \rho_{rms} C_p (r/L)^{\sigma^\rho_p}, \,\,\,\,\,\,\,\,\,\,\,\,\, 0<\sigma^\rho_p\lb{scaling3}
\end{eqnarray}
for some dimensionless constants $A_p$, $B_p$, and $C_p$. The root-mean-square of a field $f(\bx)$
is denoted by $f_{rms} \equiv\langle f^2\rangle^{1/2}$, where $\langle\dots\rangle$ is a space average.
Here, the $p$-th power of an $L_p$-norm $\| \cdot\|_p^p= \langle|\cdot|^p\rangle$ is just the traditional
structure function. 
We remark that condition (\ref{scaling3}) on the scaling of density increments
is only an upper bound. It only stipulates that the intensity of density fluctuations decays at smaller scales,
which is a very mild requirement and is readily satisfied in incompressible or nearly-incompressible
flows. Heuristically, assumptions (\ref{scaling1})-(\ref{scaling3}) characterize the roughness of 
fields: $\sigma^f <1$ specifies that the field $f(\bx)$ is ``rough enough'', while $\sigma^f >0$ states that 
$f(\bx)$ is ``smooth enough.''

Under conditions (\ref{scaling1})-(\ref{scaling3}), proving 
scale-locality of the SGS flux becomes simple and follows directly from scale-locality 
of increments \cite{Eyink05}.
For example, the contribution to any increment $\delta f(\ell)$ from scales 
$\Delta\geq\ell$ is represented by $\delta \OL{f}_\Delta(\ell)$. Here, $f(\bx)$
can denote either velocity or pressure field. Since the low-pass filtered 
field $\OL{f}_\Delta(\bx)$ is smooth, its increment may be estimated by Taylor expansion, 
and (\ref{incrementrelation1}), and (\ref{scaling1}) or (\ref{scaling2}), as
\be \|\delta \OL{f}_\Delta(\ell)\|_p \simeq \|\boell \bdot(\grad\OL{f}_\Delta)\|_p
                                                =O \left[ \left(\frac{\ell}{\Delta}\right)^{1-\sigma^f_p}\right], \lb{infraredbound}\ee
and this is negligible for $\Delta\gg\ell$ as long as $\sigma^f_p<1.$ 
The notation $O(\dots)$ denotes a big-$O$ upper bound.
On the other hand,
the contribution to any increment $\delta f(\ell)$ from scales $\delta\leq\ell$ is represented 
by $\delta f_\delta'(\ell)$. Here, $f(\bx)$ can denote either velocity or density field.
 Since $f'_\delta = {\cal O}[f(x+\delta)-f(x)]$ from (\ref{incrementrelation1}), 
 scaling conditions (\ref{scaling1}) and (\ref{scaling3}) imply that
\be \|\delta f_\delta'(\ell)\|_p \le 2\| f_\delta'\|_p 
                                        = O\left[ \left(\frac{\delta}{\ell}\right)^{\sigma^f_p}\right], \lb{ultravioletbound}\ee
and this is negligible for $\delta\ll\ell$ as long as $\sigma^f_p>0$. 
For more details and for the careful proofs of these statements, see \cite{Eyink05}
and our longer work \cite{Aluie10b}.
      
Notice that, unlike for the velocity and pressure fields, we do not stipulate that
$\rho(\bx)$ be ``rough enough.'' Contributions to the flux across scale $\ell$
from the largest density scales $L\gg \ell$ need not be negligible, yet the flux
can still be scale-local.The underlying physical reason
 is simple; an energy flux across scale $\ell$ at a point $\bx$ will depend on
 the mass in a ball of radius $\ell$ around $\bx$. Mass is proportional to average 
 density, $\OL\rho_\ell(\bx)$, in the ball which is dominated by large scales: 
 $\OL\rho_\ell(\bx)= {\cal O}[\OL\rho_L(\bx)]={\cal O}[\rho_{rms}]$.
Indeed, for incompressible turbulence,
the only density scale present is a $\bk=0$ Fourier mode, the largest possible,
and the scale-local SGS flux is directly proportional to this density mode. 
Furthermore, we do not require that the pressure field be 
``smooth enough'' even though we expect $\sigma_p^P>0$. This is because pressure only
appears as a large-scale pressure-gradient in (\ref{flux2}), with no contributions from
scales $\delta\ll \ell$.

The ultimate source of scaling properties (\ref{scaling1})-(\ref{scaling3}) 
is empirical evidence from experiments,
astronomical observations, and numerical simulations.
For incompressible turbulence, which may be viewed as a limiting case of our analysis,
the scaling of velocity and pressure structure functions (\ref{scaling1}),(\ref{scaling2}) has been 
well-established by a variety of independent studies  such as those by
\cite{Sreenivasanetal96,Xuetal07}.
Assumption (\ref{scaling3}) on density structure functions is trivially satisfied for a uniform density field.
As for compressible turbulence, the available data is also in compelling support of
our assumptions. Astronomical observations by  \cite{Armstrongetal95,Benschetal01}
show that $2$nd-order density structure function scales with 
$\sigma_2^\rho \doteq 0.3$. Measurements 
of $2$nd-order velocity structure functions in molecular clouds
\cite{Padoanetal06,HilyBlantetal08} and solar wind \cite{Salemetal09}
also yield $0<\sigma_p^u<1$ for $1\le p\le 6$.
Alongside observational evidence, numerical
studies of compressible turbulence \cite{Schmidtetal08,PriceFederrath10}
report power-law scaling exponents well within our required constraints.

Under an additional assumption concerning the co-spectrum
of pressure dilatation, which is, albeit reasonable, not as weak as 
scaling conditions (\ref{scaling1})-(\ref{scaling3}),
our proof of a scale-local SGS flux implies a scale-local \emph{conservative} cascade 
of mean kinetic energy despite the latter not being an invariant.
The requirement on pressure dilatation co-spectrum,
$E^{PD}(k)\equiv \sum_{k-0.5<|\bk|<k+0.5} -\hat{P}(\bk)\widehat{\grad\bdot\bu}(-\bk)$,
is that it decays fast enough at large $k$, 
\be |E^{PD}(k)| \le C\, u_{rms}\,P_{rms} \,(kL)^{-\beta},   \,\,\,\,\,\,\,\,\,\,\,\,\,\,\,\, \beta > 1.
\lb{scaling4}\ee
Here, $C$ is a dimensionless constant and $L$ is an integral scale.
In the limit of large Reynolds number, assumption (\ref{scaling4}) implies that
\emph{mean} pressure dilatation, $PD(\ell)\equiv \langle\OL{P}_\ell \grad\bdot\OL\bu_\ell\rangle$, 
converges to a finite constant, 
$\theta\equiv\langle P \grad\bdot\bu\rangle$, and becomes independent of $\ell$.
In other words, we have for wavenumber $K\approx \ell^{-1}$, 
\be\lim_{\ell\to 0}PD(\ell) = \lim_{K \to \infty} \sum_{0<k<K} E^{PD}(k) = \theta.
\lb{PDresult}\ee
We remark that condition (\ref{scaling4}) is sufficient but not necessary
for the convergence of $PD(\ell)$ in the limit of $\ell\to 0$. 
The series $\sum_{k<K} E^{PD}(k)$ can converge with $K \to \infty$ at a 
rate faster than what is implied by assumption (\ref{scaling4}) due to 
indefiniteness in the sign of $E^{PD}(k)$.

Saturation of mean pressure dilatation in (\ref{PDresult}) reveals 
that its role is to exchange \emph{large-scale} mean kinetic and internal energy
over a transitional ``conversion'' scale-range. At smaller scales beyond the conversion
range, mean kinetic and internal energy budgets statistically decouple.  
In other words, taking
$\ell_\mu\to 0$ first, then $\ell\to 0$,
steady-state mean kinetic energy budget becomes,
\be \langle\Pi_\ell+\Lambda_\ell \rangle  = \langle\epsilon^{inj}\rangle - \theta.
\lb{inertialKEbudget}\ee
We stress that such a decoupling is statistical and does not imply that small scales evolve
according to incompressible dynamics. Small scale compression and rarefaction can still 
take place pointwise, however, they yield a vanishing contribution to the space-average.

We denote the largest scale at which such statistical decoupling occurs by $\ell_c$. It may
be defined, for instance, as 
$\ell_c \equiv  \sum_{\bk} k^{-1} E^{PD}(\bk) / \sum_{\bk} E^{PD}(\bk)$.
Over the ensuing scale-range, $\ell_c>\ell\gg\ell_\mu$, net pressure dilatation does 
not play a role, and if, furthermore, $\langle\epsilon^{inj}\rangle$ in (\ref{inertialKEbudget})
is localized to the largest scales as shown in \cite{Aluie10a}, then
$\langle \Pi_\ell+\Lambda_\ell\rangle$ will be a constant, independent of scale $\ell$.

A constant SGS flux implies that mean kinetic energy cascades conservatively to smaller scales,
despite not being an invariant of the governing dynamics. This is one of the main conclusions
of this Letter. In particular, kinetic energy  can only reach dissipation scales via the SGS flux, 
$\Pi_\ell+\Lambda_\ell$, through a scale-local cascade process. We are therefore justified in
 calling scale-range $\ell_c>\ell\gg\ell_\mu$ the inertial range of compressible turbulence.

Needless to say, the scaling of pressure dilatation co-spectrum is easily measurable from 
numerical simulations.
We note that condition (\ref{scaling4}) does not require a 
power-law scaling  ---only that $E^{PD}(k)$ decays at a rate faster than $\sim k^{-1}$.
It is not at all trivial why one should expect $PD(\ell) = \langle\OL{P}_\ell\grad\bdot\OL\bu_\ell\rangle$ 
to converge at small scales. How can this be reconciled with the expectation that compression,
as quantified by $\grad\bdot\bu$, would get more intense at smaller scales?
Indeed, \cite*{Leeetal91} observed numerically that $(\grad\bdot\bu)_{rms}$ is an increasing
function of Reynolds number. The key point here is that our assumption (\ref{scaling4}) concerns 
\emph{spatially averaged} pressure dilatation. It is true that $\grad\bdot\bu(\bx)$, being a gradient,
derives most of its contribution from the smallest scales in the flow. Since $P\grad\bdot\bu$
is not sign-definite, however, major cancellations can occur when space-averaging.
The situation is very similar to helicity co-spectrum in incompressible turbulence, 
where the pointwise vorticity, $\bomega(\bx)=\grad\btimes\bu$, can also become unbounded in the limit
of infinite Reynolds number. Yet, numerical evidence shows that 
$\langle\bu\bdot\bomega\rangle$ remains finite and the helicity co-spectrum decays at a rate $\sim k^{-n}$,
with $n\approx 5/3 > 1$ (see for example \cite{Chenetal03} and \cite*{Kurienetal04}). 

We can offer a physical argument on why $PD(\ell)$ is expected to converge 
for $\ell\to 0$ as a result of cancellations from space-averaging.
The origin of such cancellations 
can be heuristically explained using decorrelation effects very similar to those studied
in \cite{EyinkAluie09} and \cite{AluieEyink09}. While the pressure in $\langle P\grad\bdot\bu\rangle$ derives
most of its contribution from the largest scales, $\grad\bdot\bu$ is dominated by the 
smallest scales in the flow. Therefore, pressure varies slowly in space, primarily at scales $\sim L$,
while $\grad\bdot\bu$ varies much more rapidly, primarily at scales $\ell_\mu\ll L$, leading to a 
decorrelation between the two factors. More precisely, the pressure $\OL{P}_\ell$ in 
$PD(\ell)$ may be approximated by $\OL{P}_\ell = {\cal O}[P_{rms}]= {\cal O}[\OL{P}_{L}]$ 
such that 
\begin{eqnarray}
\langle\OL{P}_\ell\grad\bdot\OL\bu_\ell\rangle 
&\approx&  \left\langle\OL{P}_L\grad\bdot\left(\OL{(\OL\bu_\ell)}_L +(\OL\bu_\ell)'_L \right)\right\rangle \nonumber\\
&\approx&  \langle\OL{P}_L\grad\bdot\OL\bu_L\rangle + \langle\OL{P}_L\rangle\langle\grad\bdot(\OL\bu_\ell)'_L\rangle. \nonumber
\end{eqnarray}
The first term in the last expression follows from $\OL{(\OL{\bu}_\ell)}_L \approx \OL\bu_L$, while
the second term is due to an approximate statistical independence between $\OL{P}_L$ and 
$\grad\bdot(\OL\bu_\ell)'_L \sim \delta u(\ell)/\ell$
which varies primarily at much smaller scales $\sim\ell \ll L$. If there is no transport beyond the 
domain boundaries or if the flow is either statistically homogeneous or isotropic, we get 
$\langle\grad\bdot(\OL\bu_\ell)'_L\rangle = 0$. The heuristic argument finally yields that pressure dilatation,
\be PD(\ell)=\langle\OL{P}_\ell\grad\bdot\OL\bu_\ell\rangle \approx \langle\OL{P}_L\grad\bdot\OL\bu_L\rangle,
\lb{PDresult2}\ee
becomes independent of $\ell$, for $\ell\ll L$. Expression (\ref{PDresult2}) corroborates our claim that the 
primary role of pressure dilatation is conversion of 
$\emph{large-scale}$ kinetic energy into internal energy and does not take part in the cascade dynamics
beyond a transitional ``conversion'' scale-range.

In summary, we conclude that there exists an inertial range in high Reynolds number  compressible turbulence
over which kinetic energy reaches dissipation scales through a conservative and scale-local cascade process.
This precludes the possibility for transfer of kinetic energy from the large-scales directly to dissipation scales,
such as into shocks, at arbitrarily high Reynolds numbers as is commonly believed. 
We make several assumptions and predictions which are amenable to empirical scrutiny. Our locality
results concerning the SGS flux can be verified in a manner very similar to what was done in \cite{AluieEyink09}
and \cite{AluieEyink10}. We also invite empirical tests of assumption (\ref{scaling4}) on
the scaling of pressure dilatation co-spectrum. Preliminary numerical results by \cite{Aluieetal11}
of compressible isotropic turbulence indicate that indeed the co-spectrum decays at a rate 
faster than $k^{-1}$. Verifying (\ref{scaling4}) or (\ref{PDresult}) under a variety of controlled conditions would substantiate the 
idea of statistical decoupling between mean kinetic and internal energy budgets. This would have potentially 
significant implications on devising reduced models of compressible turbulence.

I thank G. L. Eyink, S. S. Girimaji, S. Kurien, H. Li, S. Li, and D. Livescu.
This research was performed under the auspices of the U.S. Department of Energy at 
LANL and supported by the LANL/LDRD program.


\begin{thebibliography}{24}
\expandafter\ifx\csname natexlab\endcsname\relax\def\natexlab#1{#1}\fi
\expandafter\ifx\csname bibnamefont\endcsname\relax
  \def\bibnamefont#1{#1}\fi
\expandafter\ifx\csname bibfnamefont\endcsname\relax
  \def\bibfnamefont#1{#1}\fi
\expandafter\ifx\csname citenamefont\endcsname\relax
  \def\citenamefont#1{#1}\fi
\expandafter\ifx\csname url\endcsname\relax
  \def\url#1{\texttt{#1}}\fi
\expandafter\ifx\csname urlprefix\endcsname\relax\def\urlprefix{URL }\fi
\providecommand{\bibinfo}[2]{#2}
\providecommand{\eprint}[2][]{\url{#2}}

\bibitem[{\citenamefont{{Aluie}}(under review)}]{Aluie10a}
\bibinfo{author}{\bibfnamefont{H.}~\bibnamefont{{Aluie}}}, \bibinfo{journal}{J.
  Fluid Mech.}  (\bibinfo{year}{under review}),
  \bibinfo{note}{arXiv:1012.5877}.

\bibitem[{\citenamefont{{Kraichnan}}(1959)}]{Kraichnan59}
\bibinfo{author}{\bibfnamefont{R.~H.} \bibnamefont{{Kraichnan}}},
  \bibinfo{journal}{J. Fluid Mech.} \textbf{\bibinfo{volume}{5}},
  \bibinfo{pages}{497} (\bibinfo{year}{1959}).

\bibitem[{\citenamefont{{Eyink}}(2005)}]{Eyink05}
\bibinfo{author}{\bibfnamefont{G.~L.} \bibnamefont{{Eyink}}},
  \bibinfo{journal}{Physica D} \textbf{\bibinfo{volume}{207}},
  \bibinfo{pages}{91} (\bibinfo{year}{2005}).

\bibitem[{\citenamefont{{Domaradzki} et~al.}(2009)\citenamefont{{Domaradzki},
  {Teaca}, and {Carati}}}]{Domaradzkietal09}
\bibinfo{author}{\bibfnamefont{J.~A.} \bibnamefont{{Domaradzki}}},
  \bibinfo{author}{\bibfnamefont{B.}~\bibnamefont{{Teaca}}}, \bibnamefont{and}
  \bibinfo{author}{\bibfnamefont{D.}~\bibnamefont{{Carati}}},
  \bibinfo{journal}{Phys. Fluids} \textbf{\bibinfo{volume}{21}},
  \bibinfo{pages}{025106} (\bibinfo{year}{2009}).

\bibitem[{\citenamefont{{Aluie} and {Eyink}}(2009)}]{AluieEyink09}
\bibinfo{author}{\bibfnamefont{H.}~\bibnamefont{{Aluie}}} \bibnamefont{and}
  \bibinfo{author}{\bibfnamefont{G.}~\bibnamefont{{Eyink}}},
  \bibinfo{journal}{Phys. Fluids} \textbf{\bibinfo{volume}{21}},
  \bibinfo{pages}{115108} (\bibinfo{year}{2009}).

\bibitem[{\citenamefont{{McKee} and {Ostriker}}(2007)}]{McKeeOstriker07}
\bibinfo{author}{\bibfnamefont{C.~F.} \bibnamefont{{McKee}}} \bibnamefont{and}
  \bibinfo{author}{\bibfnamefont{E.~C.} \bibnamefont{{Ostriker}}},
  \bibinfo{journal}{Ann. Rev. Astron. Astroph.} \textbf{\bibinfo{volume}{45}},
  \bibinfo{pages}{565} (\bibinfo{year}{2007}).

\bibitem[{\citenamefont{{Germano}}(1992)}]{Germano92}
\bibinfo{author}{\bibfnamefont{M.}~\bibnamefont{{Germano}}},
  \bibinfo{journal}{J. Fluid Mech.} \textbf{\bibinfo{volume}{238}},
  \bibinfo{pages}{325} (\bibinfo{year}{1992}).

\bibitem[{\citenamefont{{Aluie} and {Eyink}}(2010)}]{AluieEyink10}
\bibinfo{author}{\bibfnamefont{H.}~\bibnamefont{{Aluie}}} \bibnamefont{and}
  \bibinfo{author}{\bibfnamefont{G.}~\bibnamefont{{Eyink}}},
  \bibinfo{journal}{Phys. Rev. Lett.} \textbf{\bibinfo{volume}{104}},
  \bibinfo{pages}{081101} (\bibinfo{year}{2010}).

\bibitem[{\citenamefont{{Aluie}}(to be published)}]{Aluie10b}
\bibinfo{author}{\bibfnamefont{H.}~\bibnamefont{{Aluie}}} (\bibinfo{year}{to be
  published}).

\bibitem[{\citenamefont{{Eyink}}(2007)}]{EyinkNotes}
\bibinfo{author}{\bibfnamefont{G.}~\bibnamefont{{Eyink}}},
  \emph{\bibinfo{title}{Turbulence theory}},
  \bibinfo{howpublished}{unpublished} (\bibinfo{year}{2007}).

\bibitem[{\citenamefont{{Sreenivasan
  $\,\,$et$\,\,$al.}}(1996)}]{Sreenivasanetal96}
\bibinfo{author}{\bibfnamefont{K.}~\bibnamefont{{Sreenivasan
  $\,\,$et$\,\,$al.}}}, \bibinfo{journal}{Phys. Rev. Lett.}
  \textbf{\bibinfo{volume}{77}}, \bibinfo{pages}{1488} (\bibinfo{year}{1996}).

\bibitem[{\citenamefont{{Xu $\,\,$et$\,\,$al.}}(2007)}]{Xuetal07}
\bibinfo{author}{\bibfnamefont{H.}~\bibnamefont{{Xu $\,\,$et$\,\,$al.}}},
  \bibinfo{journal}{Phys. Rev. Lett.} \textbf{\bibinfo{volume}{99}},
  \bibinfo{pages}{204501} (\bibinfo{year}{2007}).

\bibitem[{\citenamefont{{Armstrong} et~al.}(1995)\citenamefont{{Armstrong},
  {Rickett}, and {Spangler}}}]{Armstrongetal95}
\bibinfo{author}{\bibfnamefont{J.~W.} \bibnamefont{{Armstrong}}},
  \bibinfo{author}{\bibfnamefont{B.~J.} \bibnamefont{{Rickett}}},
  \bibnamefont{and} \bibinfo{author}{\bibfnamefont{S.~R.}
  \bibnamefont{{Spangler}}}, \bibinfo{journal}{Astrophys. J.}
  \textbf{\bibinfo{volume}{443}}, \bibinfo{pages}{209} (\bibinfo{year}{1995}).

\bibitem[{\citenamefont{{Bensch} et~al.}(2001)\citenamefont{{Bensch},
  {Stutzki}, and {Ossenkopf}}}]{Benschetal01}
\bibinfo{author}{\bibfnamefont{F.}~\bibnamefont{{Bensch}}},
  \bibinfo{author}{\bibfnamefont{J.}~\bibnamefont{{Stutzki}}},
  \bibnamefont{and}
  \bibinfo{author}{\bibfnamefont{V.}~\bibnamefont{{Ossenkopf}}},
  \bibinfo{journal}{Astron. Astroph.} \textbf{\bibinfo{volume}{366}},
  \bibinfo{pages}{636} (\bibinfo{year}{2001}).

\bibitem[{\citenamefont{{Padoan $\,\,$et$\,\,$al.}}(2006)}]{Padoanetal06}
\bibinfo{author}{\bibfnamefont{P.}~\bibnamefont{{Padoan $\,\,$et$\,\,$al.}}},
  \bibinfo{journal}{Astrophys. J. Lett.} \textbf{\bibinfo{volume}{653}},
  \bibinfo{pages}{L125} (\bibinfo{year}{2006}).

\bibitem[{\citenamefont{{Hily-Blant} et~al.}(2008)\citenamefont{{Hily-Blant},
  {Falgarone}, and {Pety}}}]{HilyBlantetal08}
\bibinfo{author}{\bibfnamefont{P.}~\bibnamefont{{Hily-Blant}}},
  \bibinfo{author}{\bibfnamefont{E.}~\bibnamefont{{Falgarone}}},
  \bibnamefont{and} \bibinfo{author}{\bibfnamefont{J.}~\bibnamefont{{Pety}}},
  \bibinfo{journal}{Astron. Astroph.} \textbf{\bibinfo{volume}{481}},
  \bibinfo{pages}{367} (\bibinfo{year}{2008}).

\bibitem[{\citenamefont{{Salem$\,\,$et$\,\,$al.}}(2009)}]{Salemetal09}
\bibinfo{author}{\bibfnamefont{C.}~\bibnamefont{{Salem$\,\,$et$\,\,$al.}}},
  \bibinfo{journal}{Astrophys. J.} \textbf{\bibinfo{volume}{702}},
  \bibinfo{pages}{537} (\bibinfo{year}{2009}).

\bibitem[{\citenamefont{{Schmidt} et~al.}(2008)\citenamefont{{Schmidt},
  {Federrath}, and {Klessen}}}]{Schmidtetal08}
\bibinfo{author}{\bibfnamefont{W.}~\bibnamefont{{Schmidt}}},
  \bibinfo{author}{\bibfnamefont{C.}~\bibnamefont{{Federrath}}},
  \bibnamefont{and}
  \bibinfo{author}{\bibfnamefont{R.}~\bibnamefont{{Klessen}}},
  \bibinfo{journal}{Phys. Rev. Lett.} \textbf{\bibinfo{volume}{101}},
  \bibinfo{pages}{194505} (\bibinfo{year}{2008}).

\bibitem[{\citenamefont{{Price} and {Federrath}}(2010)}]{PriceFederrath10}
\bibinfo{author}{\bibfnamefont{D.~J.} \bibnamefont{{Price}}} \bibnamefont{and}
  \bibinfo{author}{\bibfnamefont{C.}~\bibnamefont{{Federrath}}},
  \bibinfo{journal}{Mon. Not. R. Astron. Soc.} \textbf{\bibinfo{volume}{406}},
  \bibinfo{pages}{1659} (\bibinfo{year}{2010}).

\bibitem[{\citenamefont{{Lee} et~al.}(1991)\citenamefont{{Lee}, {Lele}, and
  {Moin}}}]{Leeetal91}
\bibinfo{author}{\bibfnamefont{S.}~\bibnamefont{{Lee}}},
  \bibinfo{author}{\bibfnamefont{S.}~\bibnamefont{{Lele}}}, \bibnamefont{and}
  \bibinfo{author}{\bibfnamefont{P.}~\bibnamefont{{Moin}}},
  \bibinfo{journal}{Phys. Fluids} \textbf{\bibinfo{volume}{3}},
  \bibinfo{pages}{657} (\bibinfo{year}{1991}).

\bibitem[{\citenamefont{{Chen $\,\,$et$\,\,$al.}}(2003)}]{Chenetal03}
\bibinfo{author}{\bibfnamefont{Q.}~\bibnamefont{{Chen $\,\,$et$\,\,$al.}}},
  \bibinfo{journal}{Phys. Rev. Lett.} \textbf{\bibinfo{volume}{90}},
  \bibinfo{pages}{214503} (\bibinfo{year}{2003}).

\bibitem[{\citenamefont{{Kurien} et~al.}(2004)\citenamefont{{Kurien}, {Taylor},
  and {Matsumoto}}}]{Kurienetal04}
\bibinfo{author}{\bibfnamefont{S.}~\bibnamefont{{Kurien}}},
  \bibinfo{author}{\bibfnamefont{M.~A.} \bibnamefont{{Taylor}}},
  \bibnamefont{and}
  \bibinfo{author}{\bibfnamefont{T.}~\bibnamefont{{Matsumoto}}},
  \bibinfo{journal}{Phys. Rev. E} \textbf{\bibinfo{volume}{69}},
  \bibinfo{pages}{066313} (\bibinfo{year}{2004}).

\bibitem[{\citenamefont{{Eyink} and {Aluie}}(2009)}]{EyinkAluie09}
\bibinfo{author}{\bibfnamefont{G.}~\bibnamefont{{Eyink}}} \bibnamefont{and}
  \bibinfo{author}{\bibfnamefont{H.}~\bibnamefont{{Aluie}}},
  \bibinfo{journal}{Phys. Fluids} \textbf{\bibinfo{volume}{21}},
  \bibinfo{pages}{115107} (\bibinfo{year}{2009}).

\bibitem[{\citenamefont{{Aluie} et~al.}(in preparation)\citenamefont{{Aluie},
  {Li}, and {Li}}}]{Aluieetal11}
\bibinfo{author}{\bibfnamefont{H.}~\bibnamefont{{Aluie}}},
  \bibinfo{author}{\bibfnamefont{S.}~\bibnamefont{{Li}}}, \bibnamefont{and}
  \bibinfo{author}{\bibfnamefont{H.}~\bibnamefont{{Li}}} (\bibinfo{year}{in
  preparation}).

\end{thebibliography}
\end{document}